\documentclass[pre,aps,twocolumn,showpacs,10pt]{revtex4-1}
\usepackage{amsmath,amssymb,graphicx}
\usepackage{epsfig}
\usepackage{textcomp}
\usepackage{color}
\usepackage{epstopdf}
\usepackage{array}
\usepackage{amssymb}
\usepackage{amsmath}
\usepackage{graphicx}
\usepackage{mathrsfs}
\usepackage[utf8]{inputenc}
\usepackage[english]{babel}
\usepackage{amsthm}
\usepackage{mathtools}

\DeclareMathAlphabet{\pazocal}{OMS}{zplm}{m}{n}
\DeclareMathOperator{\sign}{sign}
\usepackage[margin=0.5in]{geometry}
\DeclarePairedDelimiter{\ceil}{\lceil}{\rceil}

\DeclareMathOperator\arctanh{arctanh}
\DeclareMathOperator\csch{csch}
\newcommand{\HH}{\pazocal{H}}

\newcommand{\w}{\omega}

\newcommand{\la}{\lambda}
\newcommand{\kk}{\kappa}

\newcommand{\e}{\varepsilon}

\newcommand{\q}{\tilde{q}}
\newcommand{\p}{\tilde{p}}
\newcommand{\E}{\tilde{E}}
\newcommand{\T}{\pazocal{T}}

\newcommand{\pp}{\mathcal{\partial}}
\newcommand{\OO}{\mathcal{O}}

\begin{document}
\title{Population switching under a time-varying environment}
\author{Tom Israeli and Michael Assaf}
\email{michael.assaf@mail.huji.ac.il}

\affiliation{Racah Institute of Physics, Hebrew University of Jerusalem, Jerusalem 91904, Israel}

    \begin{abstract}
We examine the switching dynamics of a stochastic population subjected to a deterministically time-varying environment. Our approach is demonstrated in the realm of ecology on a problem of population establishment. Here, by assuming a constant immigration pressure along with a strong Allee effect, at the deterministic level one obtains a critical population size beyond which the system experiences establishment. Notably the latter has been shown to be strongly influenced by the interplay between demographic and environmental noise. We consider two prototypical examples for environmental variations: a temporary environmental change, and a periodically-varying environment. By employing a semi-classical approximation we compute, within exponential accuracy, the change in the establishment probability and mean establishment time of the population, due to the environmental variability. Our analytical results are verified by using a modified Gillespie algorithm which accounts for explicitly time-dependent reaction rates. Finally, our theoretical approach can also be useful in studying switching dynamics in gene regulatory networks under extrinsic variations.
\end{abstract}
\maketitle

\section{Introduction}
Stochastic populations, containing a finite number of interacting agents, typically dwell in the vicinity of some attractor, undergoing small random excursions around it. Yet, occasionally, such populations experience a rare large fluctuation, which can lead \textit{e.g.} to a transition to another attractor. Nevertheless, despite being rare, such a switching event may be of great interest in various fields such as physics, ecology, epidemiology, and biochemistry, see \textit{e.g.}~\cite{gardiner1985handbook,van1992stochastic,beissinger2002population,bartlett1960stochastic,hanski2004ecology,assaf2008noise,
meerson2008noise,shahrezaei2008analytical,escudero2009switching,assaf2011determining,biancalani2015genetic,sagi2019time}.

The problem of population switching from a long-lived metastable state is particularly important in ecology. Here, occasionally, a small population, practically on the verge of extinction, suddenly experiences a rare large fluctuation which brings it above the so-called  \textit{establishment threshold} allowing the population to establish~\cite{hanski2004ecology}. Another important example for population switching appears in the context of genetic switches governing various cellular functions; here a large fluctuation in the number of a given protein can give rise to a phenotypic switch in the cell, see \textit{e.g.}~\cite{assaf2011determining,mehta2008exponential}.

In this paper we focus on the aforementioned ecological example, and investigate how the establishment probability (EP) is influenced by a deterministically-varying environment, for example, due to a sudden temporal environmental change, or as a consequence of seasonal or diurnal effects. To undergo establishment, the underlying deterministic model must include an establishment threshold, the existence of which can be achieved, \textit{e.g.}, by incorporating the strong Allee effect. Here, the latter represents a group of phenomena in ecological models that give rise to a negative population's growth rate per capita, at population sizes below some threshold~\cite{brown1977turnover,hanski1999metapopulation,stephens1999allee,courchamp2008allee}. In this case, a population which initially resides at some pre-established (or pre-colonized) state will ultimately become established (or colonized) due to demographic noise emanating from the discreteness of individuals and stochastic nature of the interactions~\cite{brown1977turnover,hanski1999metapopulation,meerson2008noise,escudero2009switching,meerson2013immigration,be2015colonization,mendez2019demographic}.
Notably, the fact that the population initially resides at a pre-established state indicates that the population is not isolated. Otherwise, an isolated population below the establishment threshold undergoes deterministic extinction~\cite{assaf2010extinction}.

In previous studies, it has been shown that the interplay between environmental (or extrinsic) and demographic (or intrinsic) noise can dramatically affect the escape rate from a metastable state; this has been done in the context of switching between different metastable states~\cite{dykman1997resonant,dykman2001activated,elowitz2002stochastic,taniguchi2010quantifying,assaf2013extrinsic, roberts2015dynamics}, as well as in the context of extinction of stochastic populations~\cite{lande1993risks,kamenev2008colored,levine2013impact}.

Here we instead focus on the problem of population switching (or escape) under the joint effect of demographic and \textit{deterministically}-varying environment, which has  not yet been systematically studied.
To this end, we generalize a theoretical approach, initially developed to deal with extinction of stochastic populations under a time-modulated environment~\cite{assaf2008population,assaf2009population,bacaer2014probability,billings2017seasonal,vilk2018population,be2018reducing}, to population switching under time-varying environments. The analysis is based on a semi-classical approximation, in the spirit of the Wentzel-Kramers-Brillouin (WKB) method, on the pertinent master equation~\cite{Dykman1994,kessler2007extinction}, under the assumption that the typical population size is large. Doing so, in the leading order this analysis yields a Hamilton-Jacobi equation with an explicitly time-dependent Hamiltonian, which governs the system's dynamics, and from which the EP can be computed within exponential accuracy. Our approach is demonstrated on a problem of population establishment, using a variant of the well-known Verhulst model, where we account for external variations in two ways: (i) a temporary change in the environment, and (ii) a time-periodic environment. We furthermore include analysis of both additive as well as multiplicative variability. Finally, to examine the validity of our theoretical results, we implement a modified version of the Gillespie algorithm~\cite{gillespie1976general} which accounts for explicitly time-dependent reaction rates~\cite{anderson2007modified}.

The paper is organized as follows. In Sec.~\ref{Deterministic} we analyze the deterministic model, while in Sec.~\ref{Stochastic} we include demographic noise, and use the WKB method to find the EP and mean time to establishment (MTE) in the absence of external perturbation. In Sec~\ref{PerturbedStochastic} we apply various perturbation techniques and find the change in the EP and MTE under two different protocols of environmental variations, see below. Our numerical algorithm is described in Sec.~\ref{SimulationMethods}, while a summary and discussion are given in Sec.~\ref{summary}.

\section{Theoretical Model and deterministic analysis} \label{Deterministic}
We consider a variant of the Verhulst logistic model which includes the Allee effect and constant immigration flux. The model can be described by the following birth-death reactions and corresponding rates:
\begin{eqnarray} \label{reactions}
&&n\xrightarrow[]{N\lambda(n)}n+1\;;\;\;\;\;
\lambda(n) = f_0+ \frac{n^2}{2N^2},\; \nonumber\\
&&n\xrightarrow[]{N\mu(n)}n-1\;;\hspace{4mm}\mu(n)=n/N,
\end{eqnarray}
where $n$ is the population size, $N\gg 1$ is the typical population size prior to switching and $0<f_0<1/2$ measures the flux magnitude of individuals entering the system.

Ignoring fluctuations, and defining the population density, $q=n/N$, the deterministic dynamics of the mean population density $\bar{q}$, is governed by the following rate equation:
\begin{equation} \label{eq:rate_eqn}
\frac{d\bar{q}}{dt} = \lambda(\bar{q}) - \mu(\bar{q}) = f_0 + \frac{\bar{q}^2}{2} - \bar{q},
\end{equation}
where $\lambda(q)=\lambda(n/N)$ and $\mu(q)=\mu(n/N)$. This equation has an attractive fixed points at $q_1 = 1-\delta$ and a repelling fixed point at $q_2=1+\delta$, where $\delta = \sqrt{1-2f_0}>0$. The characteristic relaxation time of the system in the vicinity of $q_1$ is $t_r = 1/\delta$. In the case of time-\textit{independent} reaction rates, Eq.~(\ref{eq:rate_eqn}) can be exactly solved, yielding
\begin{equation}\label{ratesol}
\bar{q}(t) =
1-\delta \tanh\left[\delta(t-t_0)/2\right].
\end{equation}
Note, that in our model the threshold for establishment is given by $n_2=N(1+\delta)$, while for simplicity, we have taken the established state to be at infinity. As a result, the switching problem, see below, becomes an effective problem of noise-driven population explosion~\footnote{In order to have a finite established state, a nonlinear cubic term has to be added to the death rate. However, if the population size at the established state is sufficiently large~\cite{be2015colonization}, this term has no effect in the leading order on the MTE.}.

\begin{figure}[t]
	\includegraphics[width=0.95\linewidth]{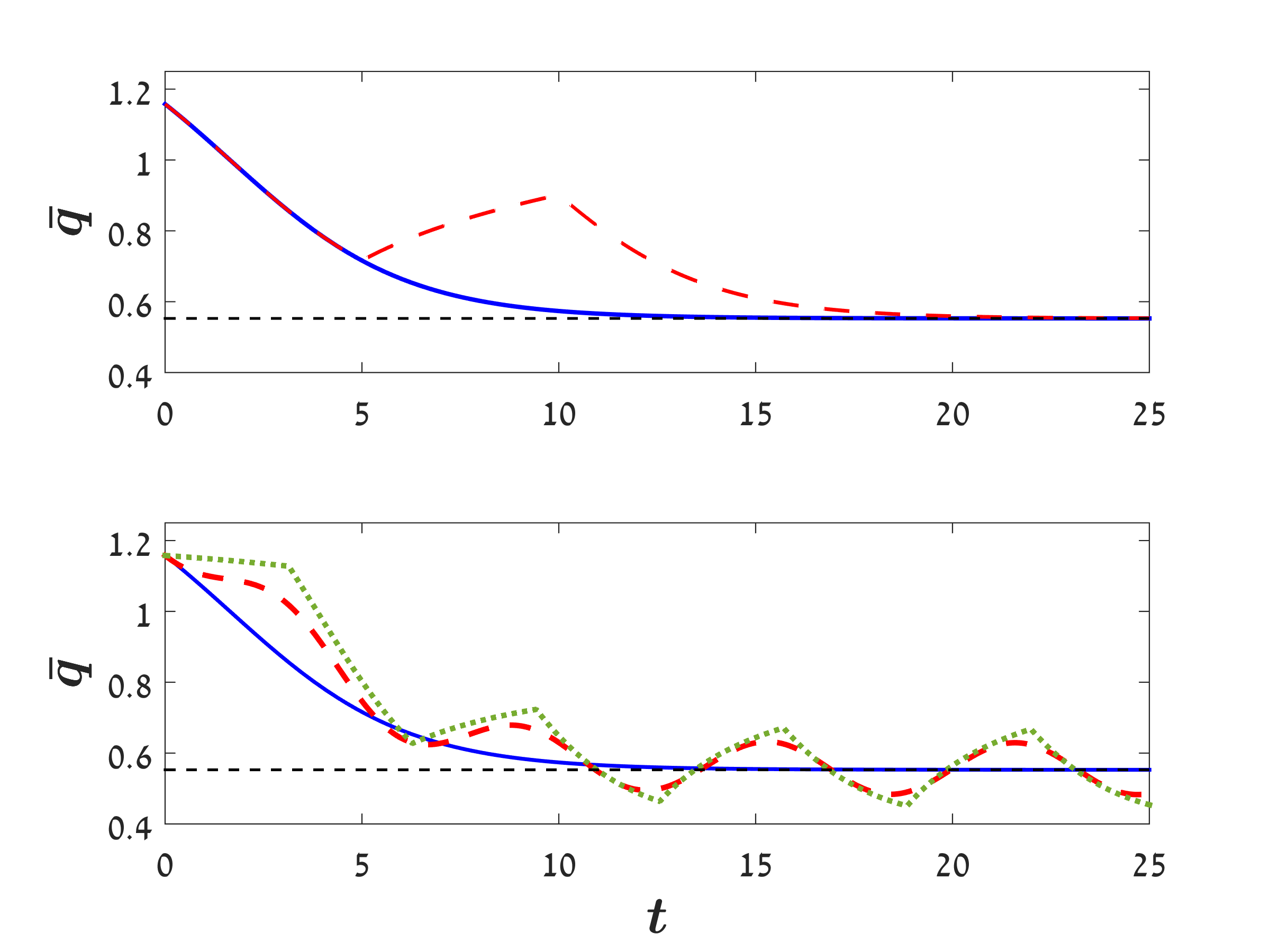}
	\caption{A comparison between the solution of equation~(\ref{eq:rate_eqn}) in the unperturbed and perturbed cases. In the latter, the incoming flux becomes $f=f_0+\phi(t)$, see Sec.~\ref{PerturbedStochastic}.
Upper panel: the unperturbed solution (solid), given by Eq.~(\ref{ratesol}), is compared with a temporarily-perturbed flux, with $\phi=0.12$ at times $5\leq t\leq 10$ and $\phi=0$ otherwise. Lower panel: the unperturbed solution (solid line), given by Eq.~(\ref{ratesol}), is compared with two examples of a time-periodic flux: a sinusoidal flux $f(t)=f_0[1+\e\sin(\w t)]$ (dashed red line), and a square wave perturbation $f(t)=f_0\{1+\e\sign\left[\sin(\w t)\right]\}$ (dotted green line). The parameters here are $f_0=0.4$, $\w=1$ and $\e=0.2$. In both panels dashed horizontal line correspond to the attractive (pre-established) fixed point $q_1=1-\delta$.}
	\label{fig1}
\end{figure}

An example of the solutions to Eq.~(\ref{eq:rate_eqn}) in the cases of constant and time-perturbed environment is shown in Fig.~\ref{fig1}. One can see that the solutions in the case of a perturbed environment follow a trend given by Eq.~(\ref{ratesol}). In the following, we consider two time-dependent scenarios: additive perturbation in the birth rate, and multiplicative perturbation in the death rate, see Sec.~\ref{PerturbedStochastic}.

\section{Stochastic Analysis of the Unperturbed Case} \label{Stochastic}
Here we derive the results for the mean time to establishment (MTE) and establishment probability (EP) in the case of constant environment, \textit{i.e.}, the unperturbed case, following Refs.~\cite{meerson2008noise,escudero2009switching}. Accounting for demographic noise, the attractive fixed point in the language of the deterministic rate equation, $n_1 = N(1-\delta)$, becomes metastable, and escape over the barrier at $n_2 = N(1+\delta)$ eventually occurs with unit probability, see below. To find the MTE, we write down the master equation describing the evolution of the probability $P(n,t)$ to have $n$ individuals at time $t$:
\begin{eqnarray} \label{MasterEquation}
\frac{\pp P(n,t)}{\pp t} &=&
\left[ Nf_0 + \frac{(n\!-\!1)^2}{2N} \right] P(n\!-\!1,t) +
(n\!+\!1)P(n\!+\!1,t)\nonumber\\
&-&\left[Nf_0 + \frac{n^2}{2N} + n \right]P(n,t).
\end{eqnarray}
For $N\gg 1$, after a short time transient on the order of $t_r$, the system enters the long-lived metastable state centered about $n_1$. This metastable state slowly decays in time due to an infinitesimally small probability flux through the unstable fixed point $n=n_2$, which approximately equals the inverse of the MTE. As a result, at $t\gg t_r$, one can write $P(n\leq n_2,t)\simeq \pi(n)e^{-t/\tau}$, while $\sum_{n>n_2}P(n,t)=1-e^{-t/\tau}$. Here, $\pi(n)$ is the (normalized) quasi-stationary distribution (QSD), centered about $n=n_1$ and represents the shape of the metastable state, and $\tau$ is the MTE~\cite{Dykman1994,meerson2008noise,escudero2009switching,assaf2017wkb}. To find the MTE, we employ the WKB ansatz, $\pi(n)=A\exp[-NS(q,t)]$ in the (quasi)stationary master equation [Eq.~(\ref{MasterEquation}) with $\partial_t P(n,t)=0$], where $S(q,t)$ is the action function in terms of the normalized coordinate $q=n/N$, and $A$ is an unknown prefactor. Doing so, and keeping only leading-order terms in $N\gg 1$, we arrive at a stationary Hamilton-Jacobi equation, $\HH_0 (q,p)=0$, with
\begin{equation} \label{ZeroHamiltonian}
\HH_0 (q,p) = \left[ f_0+\frac{q^2}{2} -qe^{-p} \right](e^p-1),
\end{equation}
where $p=\pp S / \pp q$ plays the role of the momentum conjugate to the coordinate $q$. Note, that here, the Hamiltonian does not depend on time explicitly, and thus it is an integral of motion. Thus,
to find the optimal path to switch -- the path the system takes with an overwhelmingly large probability during a switching event -- we need to find a nontrivial heteroclinic trajectory, $p_0(q)$, connecting the saddles $(q,p) = (q_1,0)$ and $(q_2,0)$~\cite{dykman1979theory,moss1989noise,Dykman1994,freidlin1998random}. Equating $\HH_0=0$ yields
\begin{equation} \label{eq:unperturbedOptimalPath}
p_0(q) = \ln\left[ \frac{2q}{1-\delta^2 + q^2} \right].
\end{equation}
Indeed, this path leaves, at $t=-\infty$, the saddle point $(q,p)=(1-\delta,0)$ along its unstable manifold, and arrives, at $t=\infty$, at the saddle point $(q,p)=(1+\delta, 0)$. Alternatively, the optimal path can be found by solving the Hamilton equations:
\begin{eqnarray}
 \label{HamiltonEquations}
\dot{q} &=& \frac{\pp \HH}{\pp p} = \left[f_0 + \frac{q^2}{2}\right] e^p - q e^{-p},\nonumber\\
\dot{p} &=& -\frac{\pp \HH}{\pp q} = 1 - e^{-p} + q - q e^p.
\end{eqnarray}
The solution of these equations, with initial conditions $q(t=-\infty)=1-\delta$ and $p(t=-\infty)=0$, is given by:
\begin{eqnarray} \label{eq:ZeroTDtajectories}
q_0(t-t_0) &=& \sigma(t-t_0)\;;\;\;\;
\sigma(t) = 1-\delta \tanh\left(\frac{\delta t}{2}\right),
\nonumber\\
p_0(t-t_0) &=& \ln\left[ \frac{ 2\sigma(t-t_0)}{1-\delta^2+\sigma^2(t-t_0)} \right],
\end{eqnarray}
where $t_0$ is an arbitrary time shift. As a result, using Eq.~(\ref{eq:unperturbedOptimalPath}) or Eqs.~(\ref{eq:ZeroTDtajectories}), the action along the optimal path, $S_0=\intop_{-\infty}^{\infty}p_0(t)\dot{q_0}(t)dt=\intop_{q_1}^{q_2} p_0(q)dq$, becomes
\begin{equation} \label{eq:ZeroAction}
S_0 = \intop_{1-\delta}^{1+\delta}\ln\left[ \frac{2q}{1-\delta^2 + q^2} \right]dq
= 2\delta - 2\sqrt{1-\delta^2}\arcsin(\delta).
\end{equation}
Finally, since the probability flux through $n_2$ is proportional to $e^{-NS_0}$, we find the MTE to be~\cite{meerson2008noise,escudero2009switching}:
\begin{equation} \label{eq:ZeroEstablishmentRate}
\tau = \frac{2\pi\sqrt{1+\delta}}{\delta\sqrt{1-\delta}} e^{N S_0},
\end{equation}
where the pre-exponential factor has been found using reaction rates~(\ref{reactions}) and Eq.~(23) in Ref.~\cite{escudero2009switching}. In addition to computing $\tau$, one can also calculate the time-dependent EP, $\mathcal{P}(t)$ -- the probability that the system undergoes switching up to time $t$. In the case of exponentially-long $\tau$, the latter is given by $\mathcal{P}(t)=\sum_{n>n_2}P(n,t)\simeq 1-e^{-t/\tau}$~\cite{meerson2008noise}. As a result, at not too long times $t\ll \tau$, we have $\mathcal{P}\simeq t/\tau\sim \tau^{-1}\sim e^{-NS_0}$.

\section{Stochastic Analysis of the Perturbed case} \label{PerturbedStochastic}
In this section we incorporate a time-varying environment into the model, by considering two scenarios of explicitly time-dependent reaction rates: additive variation in the birth rate, and multiplicative variation in the death rate:
\begin{eqnarray}
\hspace{-6mm}\lambda(q,t) &=& f_0 + \phi(t) + \frac{q^2}{2},\;\;\mu(q,t) = q,\;\;\mbox{(additive)}\label{eq:reactionsAdditive}\\
\hspace{-6mm}\lambda(q,t) &=& f_0 + \frac{q^2}{2},\;\;\mu(q,t) = q[1-\phi(t)],\;\;\mbox{(multiplicative)}\label{eq:reactionsMultiplicativeConst}
\end{eqnarray}
where $\phi(t)$ represents the environmental perturbation.
In order to calculate the EP and MTE in these cases, one can repeat the semi-classical treatment done in the previous section and arrive at a Hamilton-Jacobi equation, $\pp S/\pp t = -\HH(q,p,t)$, with an explicitly time-dependent Hamiltonian:
\begin{equation} \label{timeDependentHamiltonian}
\HH(q,p,t)=\left[ \lambda(q,t)-\mu(q,t)e^{-p} \right](e^p-1),
\end{equation}
with the reaction rates given by Eq.~(\ref{eq:reactionsAdditive}) or~(\ref{eq:reactionsMultiplicativeConst}).

As in the case of constant (unperturbed) environment, the action $S$ can be computed by integrating along the optimal path to switch $\{q_{op}(t),p_{op}(t)\}$. Yet, since now the Hamiltonian explicitly depends on time, it is no longer an integral of motion. Thus, as in general $dS = (\pp S/\pp t) dt + (\pp S /\pp q) dq$, the action along this heteroclinic trajectory satisfies~\cite{assaf2009population}:
\begin{equation} \label{ActionFromOptimalPath}
S = \intop_{-\infty}^{\infty}
\left\{ p_{op}(t)\dot{q}_{op}(t) - \HH\left[ q_{op}(t),p_{op}(t),t \right] \right\}dt.
\end{equation}
Having found the action associated with the time-dependent rates, the EP and MTE can be found, as detailed below.

\subsection{Temporary perturbation}
In this subsection we examine the case of a temporal change in the environment, where at some arbitrary time $t_0$, the environment switches to a new (constant) state, for a finite period of time $T$, whereas at time $t_0+T$ the system transitions back to the original environmental state. Here, we assume that $\phi>0$ in Eqs.~(\ref{eq:reactionsAdditive})-(\ref{eq:reactionsMultiplicativeConst}); that is, the new environmental state is advantageous for population establishment, as respectively either the incoming flux increases or death rate decreases due to the perturbation.

The question we address here is how the EP increases due to this environmental change occurring at $t=t_0$. In fact, at time $t<t_0$, but much shorter than the MTE, the EP is exponentially small, and is given by $\mathcal{P}_{BP}\simeq 1-e^{-t/\tau}\simeq t/\tau$, where $\tau$ is the MTE~\cite{meerson2008noise}, see end of Sec~\ref{Stochastic}. At times $t_0<t<t_0+T$, when the environment becomes beneficial for establishment, the EP grows at a faster rate, reaching a value of $\mathcal{P}_{AP}$ at $t=t_0+T$, which satisfies $\mathcal{P}_{AP}\gg \mathcal{P}_{BP}$. In the following we compute $\Delta \mathcal{P}\equiv\mathcal{P}_{AP}-\mathcal{P}_{BP}\simeq \mathcal{P}_{AP}$ -- the \textit{increase} in the EP due to the environmental perturbation -- by employing the WKB approximation~\cite{assaf2009population}:
\begin{equation}\label{DP}
\Delta \mathcal{P} \sim e^{-NS},
\end{equation}
where $S$ is the action in the aftermath of the environmental change~\footnote{In this scenario, while $\Delta \mathcal{P}$ turns out to be significant, the MTE is almost unaffected by the temporary perturbation, as its duration is finite, and much shorter than the MTE.}. Note, that our treatment below extends the results in~\cite{assaf2009population}; here an environmental change of arbitrary magnitude and duration is considered, while in~\cite{assaf2009population} the analysis was carried out by presetting the perturbation magnitude.

As stated before, a change to a beneficial environment can occur by either increasing the birth rate, $f_0\to f_0+\phi$, or by decreasing the death rate, $\mu\to \mu(1-\phi)$, see Eqs.~(\ref{eq:reactionsAdditive})-(\ref{eq:reactionsMultiplicativeConst}). In the case of a temporary environmental change, $\phi$ satisfies
\begin{equation} \label{eq:perturbationConstant}
\phi(t)=\begin{cases}
0, & t<t_0,\;\;t>t_0+T \\
F, & t_0\le t \le t_0+T.
\end{cases}
\end{equation}
Here, $F$ can receive any positive value in the additive case, while in the multiplicative case, one must have $F<1$.

The action $S$ in the aftermath of the environmental change can be computed by integrating along the optimal path to switch, using Eq.~(\ref{ActionFromOptimalPath}). This heteroclinic trajectory starts at the saddle point $(q,p)=(q_1,0)$ well before the perturbation has been applied, and ends at the saddle point $(q,p)=(q_2,0)$ well after the perturbation has been terminated.

\begin{figure}[t]
	\includegraphics[width=0.81\linewidth]{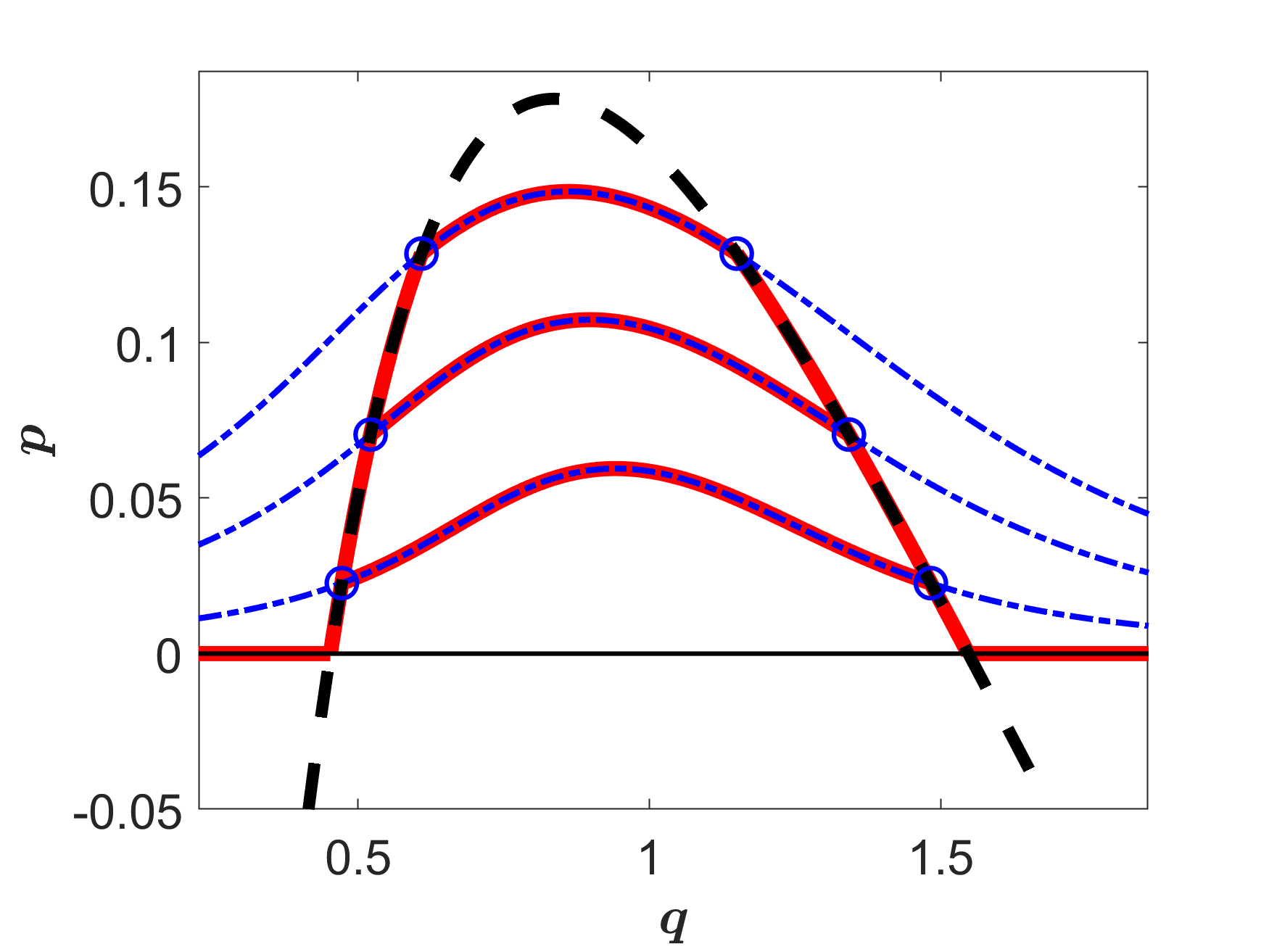}
	\caption{Phase space trajectories. The unperturbed trajectory $p_0(q)$ [Eq.~(\ref{eq:unperturbedOptimalPath})] (dashed line), three perturbed constant-energy trajectories, $p_p(q;E_p)$, for an additive perturbation [Eq.~(\ref{eq:PullingTrajectory}) with reaction rates given by~(\ref{eq:reactionsAdditive})] (crushed lines) and the predicted optimal path to switch (thick solid line) for each perturbation. The parameters are $f_0=0.35$, $F=0.15$ and $T=\{2,4,10\}$ (for which $E_p=\{0.021,0.011,0.003\}$) for the perturbed trajectories, from top to bottom. The blue open circles are the intersection points between the unperturbed and perturbed trajectories, $q_1^p(E_p)$ (left circles) and $q_2^p(E_p)$ (right circles). As the perturbation duration increases, apart from the $E_pT$ term, see Eq.~(\ref{eq:actionConstForce}), the action decreases by an amount which equals the area between the dashed and thick solid lines. The same qualitative behavior is obtained when $T$ is kept constant and $F$ is increased.}
	\label{fig2}
\end{figure}

It turns out that in this case, despite having a time-varying environment, the optimal path to switch can be analytically found. This is because in this scenario the environment changes from one constant value to another, and thus there are now two different time-independent Hamiltonian functions, both of which are integrals of motion. These are the unperturbed Hamiltonian at times $t<t_0$ and $t>t_0+T$, $\HH(q,p)=\HH_0(q,p)$, see Eq.~(\ref{ZeroHamiltonian}), and the perturbed Hamiltonian, $\HH(q,p)=\HH_p(q,p)$, at times $t_0<t<t_0+T$,
where
\begin{equation} \label{PullingHamiltonian}
\HH_p (q,p) = \left[ \lambda_p(q) -\mu_p(q)e^{-p} \right](e^p-1).
\end{equation}
Here, $\lambda_p$ and $\mu_p$ are the perturbed reaction rates given by Eqs.~(\ref{eq:reactionsAdditive}) or (\ref{eq:reactionsMultiplicativeConst}) with $\phi(t)=F$ [see Eq.~(\ref{timeDependentHamiltonian})].
Yet, while the optimal path before and after the perturbation, $p_0(q)$, is determined by the zero-energy trajectory of~(\ref{eq:unperturbedOptimalPath}), $\HH_0(q,p)=0$, during the perturbation, the energy equals some $E_p$, and is no longer zero. Solving $\HH_p(q,p_p)=E_p$ for the \textit{perturbed} trajectory $p_p(q)$, where $E_p$ is a-priori unknown, we find:
\begin{eqnarray} \label{eq:PullingTrajectory}
\hspace{-4mm}p_p(q;E_p) \!=\! \ln\! \left[\frac{\la_p\!+\!\mu_p\!+\!E_p\!+\!\sqrt{(\la_p\!+\!\mu_p\!+\!E_p)^2\!-\!4\mu_p\la_p}}{2\la_p}\right]\!\!.
\end{eqnarray}
The energy $E_p$ is implicitly determined by demanding that the duration of the perturbed trajectory be exactly $T$:
\begin{equation} \label{T_Ep_relation}
T=\intop_0^T dt = \intop_{q_1^p (E_p)}^{q_2^p (E_p)} \frac{dq}{\dot{q}\left[q,p_p(q;E_p)\right]},
\end{equation}
where the integral boundaries, $q_1^p (E_p)$ and $q_2^p (E_p)$, are the intersection points between the unperturbed optimal path $p_0(q)$, given by Eq.~(\ref{eq:unperturbedOptimalPath}), and the perturbed trajectory $p_p(q;E_p)$, given by Eq.~(\ref{eq:PullingTrajectory}). In Fig.~\ref{fig2} we demonstrate these intersections by plotting an example of the optimal path to switch for different perturbation durations.

To explicitly compute $E_p$, we substitute Hamilton's equation for $\dot{q}$ along the perturbed path, $\dot{q}=\partial_p \HH_p(q,p)=\lambda_p(q)e^p-\mu_p(q)e^{-p}$, evaluated at $p=p_p(q;E_p)$ [see Eq.~(\ref{eq:PullingTrajectory})], into Eq.~(\ref{T_Ep_relation}), and arrive at an algebraic equation for $E_p$:
\begin{equation} \label{eq:Tcondition}
T = \intop_{q_1^p (E_p)}^{q_2^p (E_p)} \frac{dq}{\la_p(q)e^{p_p(q;E_p)}-\mu_p(q)e^{-p_p(q;E_p)}}.
\end{equation}
\begin{figure}[t]
	\includegraphics[width=0.95\linewidth]{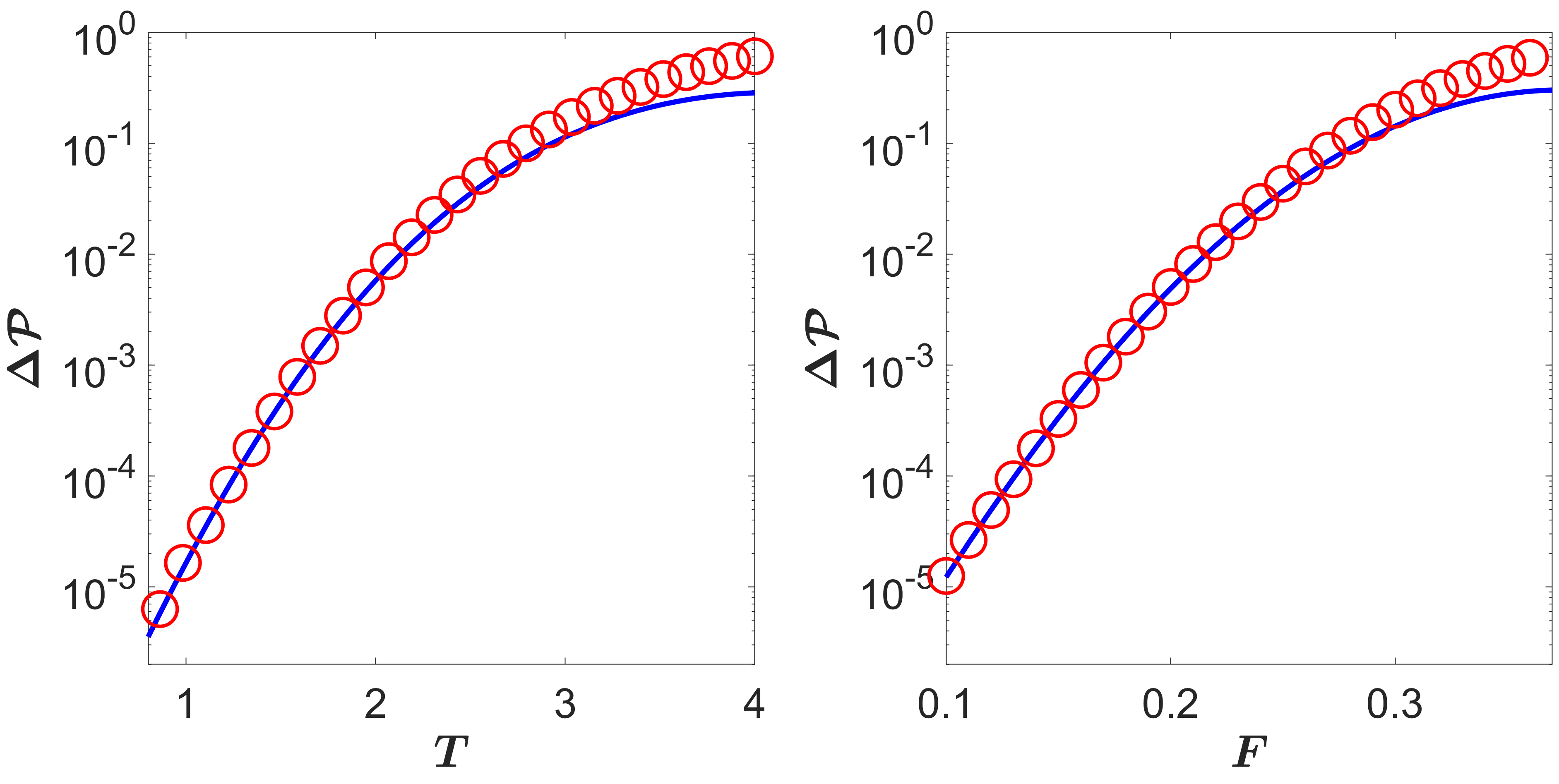}
	\caption{The change in the establishment probability (EP), $\Delta\mathcal{P}$, in the aftermath of a temporary environmental change. Left panel: $\Delta\mathcal{P}$ as function of perturbation duration T, with $F=0.25$. Right panel: $\Delta\mathcal{P}$ as function of perturbation magnitude F, with $T=2.5$. In both panels the theoretical result~(\ref{DP}) with~(\ref{eq:actionConstForce}) (solid line) is compared with Monte-Carlo simulations (symbols), where the theoretical result is multiplied by a constant prefactor of 0.3 in order to enable direct comparison. Parameters are $N=400$ and $f_0=0.42$. The disagreement at high $F$ and $T$ occurs as the action becomes  $\OO (1)$  and the WKB approximation breaks down.}
	\label{fig3}
\end{figure}
Finally, the action can be computed using Eq.~(\ref{ActionFromOptimalPath}):
\begin{eqnarray} \label{eq:actionConstForce}
S &=& \intop_{q_1}^{q_2} p_0(q)dq - \intop_{q_1^{p}(E_p)}^{q_2^{p}(E_p)} \big[p_0(q) - p_p(q;E_p) \big]dq \!-\! \intop_{t_0}^{t_0+T} \HH_p dt \nonumber\\
&=& S_0 - E_p T - \intop_{q_1^{p}(E_p)}^{q_2^{p}(E_p)} \big[p_0(q) - p_p(q;E_p) \big]dq,
\end{eqnarray}
where $S_0$ is the unperturbed action from Eq.~(\ref{eq:ZeroAction}), while $E_p=E_p(F,T)$ is found from Eq.~(\ref{eq:Tcondition}).

In Fig.~\ref{fig3} we compare the theoretical result for the change in the EP, $\Delta\mathcal{P}$ [Eqs.~(\ref{DP}) and (\ref{eq:actionConstForce})], with numerical Monte-Carlo simulations, see Sec.~\ref{SimulationMethods}, in the aftermath of the environmental perturbation. Here, we plot $\Delta\mathcal{P}$ as a function of the perturbation magnitude $F$, and duration $T$, for the case of additive perturbation [see Eq.~(\ref{eq:reactionsAdditive})]. One can see that increasing either $F$ or $T$ results in a decrease of the action (see also Fig.~\ref{fig2}) and an increase in $\Delta\mathcal{P}$. Similar results (not shown) are obtained for the case of multiplicative perturbation, as long as $F$ is not too close to $1$, see below.

Note, that in Fig.~\ref{fig3}, as well as Figs.~\ref{fig4} and~\ref{fig5}, we have multiplied the theoretical result by a constant prefactor to match the results of the simulations in their joint region of applicability. This is because our theoretical results (apart from the adiabatic case, see below) are obtained only within exponential accuracy.

\subsubsection{\textbf{Bifurcation Limit}} \label{ConstantForceBifurcationLimit}
The above results drastically simplify close to the bifurcation limit, for $\delta\ll 1$, where the stable and unstable fixed points, $q_1=1- \delta$ and $q_2=1+\delta$, become close. Here, for simplicity, we only consider the case of additive perturbation, whereas the multiplicative case can be treated in a similar manner. In the following, it is convenient to use shifted and rescaled coordinate and momentum
\begin{equation} \label{RescaledCoordinates}
\q=\frac{q-1}{\delta}\;;\;\,
\p = \frac{p}{\delta^2},
\end{equation}
where $\q,\p\sim\OO(1)$, see below. With this definition of $\q$ and $\p$, in the leading order in $\delta\ll 1$, the unperturbed optimal path~(\ref{eq:unperturbedOptimalPath}) becomes a parabola: $\p_0(\q)=(1-\q^2)/2 + \OO(\delta)$.
Expanding perturbed Hamiltonian~(\ref{PullingHamiltonian}) in $\delta\ll 1$, we find
\begin{equation}\label{perthambif}
\HH_p(\q,\p) = F\p\,\delta^2 + \OO(\delta^4),
\end{equation}
where we have assumed $F=\OO(1)$. Equating $\HH_p = E_p$, and defining $\tilde{E}_p=E_p/\delta^2=\OO(1)$, see below, the perturbed optimal path reads $\p_p(\q) = \tilde{E}_p/F + \OO(\delta^2)$, which is almost constant close to bifurcation. The intersection points are found by solving $\p_p(\q)=\p_0(\q)$, which yields $\q_{1}^p (\E_p) = -(1-2\E_p/F)^{1/2}$ and $\q_{2}^p (\E_p) = (1-2\E_p/F)^{1/2}$, while $\E_p$ can be found from Eq.~(\ref{T_Ep_relation}), by computing $\dot{q}$ along the perturbed path. Indeed, differentiating the perturbed Hamiltonian~(\ref{perthambif}) with respect to $p$ and using Eq.~(\ref{RescaledCoordinates}), we find $\dot{q}\left[q,p_p(q;E_p)\right] = F + \OO(\delta^2)$.
Plugging this into~(\ref{T_Ep_relation}) we find $T= (2\delta/F) (1-2 \E_p/F)^{1/2}$, which yields the rescaled energy
\begin{equation} \label{EpBifurcation}
\E_p(F,\T) = (F/2)\left[1 - \left(F\T/2\right)^2 \right],
\end{equation}
where $\T = T/ \delta=\OO(1)$. This result is valid as long as $F\T<2$ or $T<2\delta/F$. This condition stems from the fact that when the system is close to bifurcation, for a long duration $T = \OO(1)$, even a very small perturbation magnitude, $F = \OO(\delta)$, is sufficient to give rise to a deterministic establishment of the population, and to cause a significant increase in the EP. Alternatively, for $F = \OO(1)$, even a short perturbation $T = \OO(\delta)$ is sufficient to significantly increase the EP.

Finally, after some algebra, the action [Eq.~(\ref{eq:actionConstForce})] close to the bifurcation limit, reads
\begin{equation} \label{eq:dS_bifurcation}
S = \frac{2}{3} \delta^3
\left\{1 - \frac{3}{4}F\T  \left[1 - \frac{(F\T)^2}{12}\right] \right\}.
\end{equation}
where $(2/3)\delta^3$ corresponds to the unperturbed action, $S_0$, in the leading order in $\delta\ll 1$~\cite{Dykman1994}.

\begin{figure}[t]
	\includegraphics[width=0.81\linewidth]{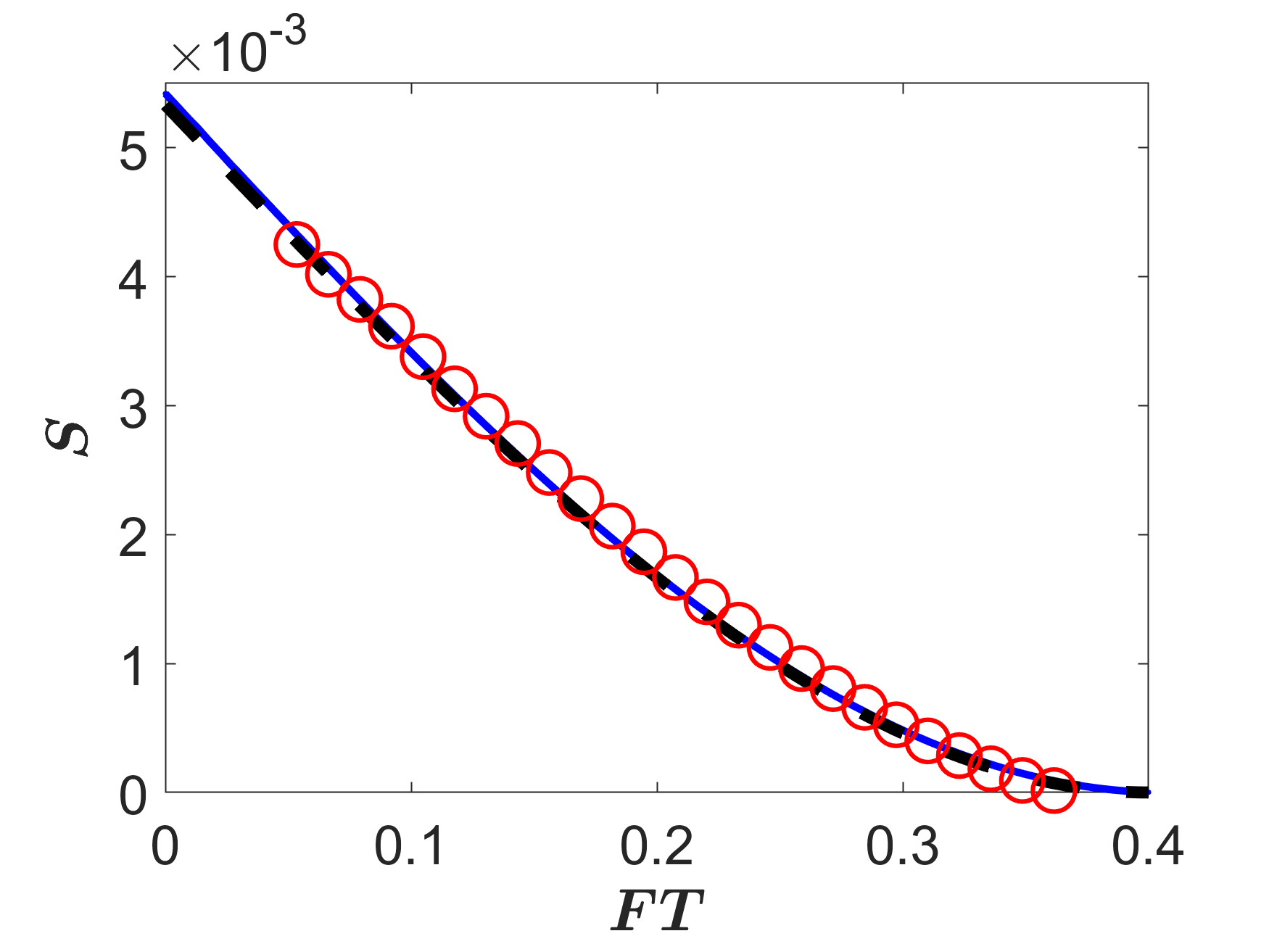}
	\caption{The action $S$ as function of the product $FT$ close to the bifurcation limit. Here we compare between the bifurcation result~(\ref{eq:dS_bifurcation}) (dashed line), the full analytical solution~(\ref{eq:actionConstForce}), and numerical Monte-Carlo simulations with $S=-\log(\Delta\mathcal{P})/N$ (symbols). The parameters are $N=3000$, $\delta=0.2$, $T=0.15$ and $0\le F\le 2.7$, while $\Delta\mathcal{P}$ is multiplied by a constant prefactor of 2.}
	\label{fig4}
\end{figure}

As stated before, a necessary condition for Eq.~(\ref{eq:dS_bifurcation}) to be valid is that $F\T<2$; otherwise $S$ vanishes and becomes negative. Interestingly, Eq.~(\ref{eq:dS_bifurcation}) shows that close to bifurcation, the action and EP depend only on the product $FT$ and not on $F$ and $T$ separately. This behavior is demonstrated in Fig.~\ref{fig4}, where the analytical expression close to bifurcation~(\ref{eq:dS_bifurcation}) is shown to agree well with  Monte-Carlo simulations, and with the full solution given by Eq.~(\ref{eq:actionConstForce}) with condition (\ref{eq:Tcondition}).

\subsection{Periodic perturbation} \label{Sinusoidal}
In this subsection we consider a different environmental perturbation which is not temporary but of infinite duration. Here we take a time-periodic perturbation with a given amplitude and frequency. For concreteness, we consider a sinusoidal perturbation added to the constant flux such that:
\begin{equation} \label{eq:sinusoidal_pert}
f(t) = f_0 \left[1 + \e\sin (\w t) \right],
\end{equation}
where $\w=2\pi/T$ is the angular frequency of the perturbation, and $\e$ is the amplitude. The resulting Hamiltonian can be written as
\begin{equation} \label{eq:H0plusH1}
\HH(q,p;t) = \HH_0(q,p) + \varepsilon\HH_1(q,p;t),
\end{equation}
where $\HH_0$ is given by unperturbed Hamiltonian~(\ref{ZeroHamiltonian}) and
\begin{equation}
\label{eq:H1}
\HH_1(q,p;t) = (e^p-1) f_0 \sin(\w t).
\end{equation}

In order to compute the effect the time-periodic environment has on the MTE, one needs to calculate the action along the optimal path; the latter is now time-dependent, and is denoted by $\{q(t,t_0),p(t,t_0)\}$. Indeed, using Hamiltonian~(\ref{eq:H0plusH1}), the action [Eq.~(\ref{ActionFromOptimalPath})] becomes~\cite{assaf2008population}:
\begin{eqnarray} \label{eq:Sperturbed}
S &=& \intop_{-\infty}^{\infty}
\big\{
p(t,t_0)\dot{q}(t,t_0) - \HH_0\left[ q(t,t_0),p(t,t_0),t \right]
\nonumber\\
&-&\varepsilon \HH_1[q(t,t_0),p(t,t_0),t]
\big\}dt,
\end{eqnarray}
where $\dot{q}(t,t_0)=dq/dt$ and $\HH_0(q,p)$ is invariant to the specific choice of $t_0\in [0,T]$. In the following we show that $t_0$ is determined in such a way to minimize the action~\cite{dykman1997resonant,dykman2001activated,escudero2008persistence}.

In the next two subsections we find the action in two important limits: weak periodic perturbation, and slowly-varying (or adiabatic) perturbation. In the latter case, the perturbation frequency $\omega$ is much smaller than the system's relaxation rate $t_r^{-1}\sim\delta$, such that $\omega\ll \delta$.
In the opposite limit of rapidly-varying perturbation, $\w\gg\delta$, the MTE is almost unaffected by the periodic environment~\cite{assaf2008population}, and thus we will not deal with this case here.

\begin{figure}[t]
	\includegraphics[width=0.81\linewidth]{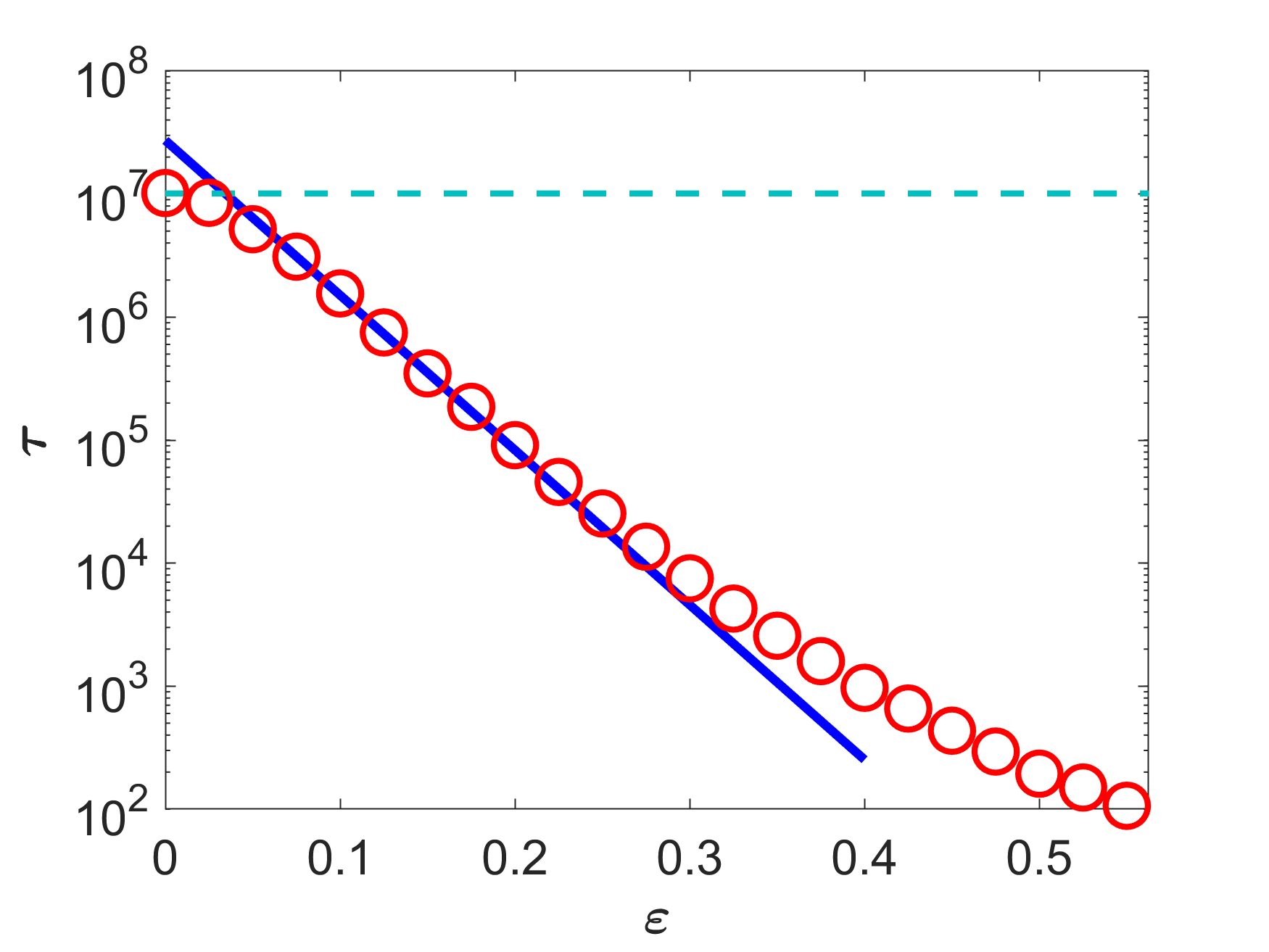}
	\caption{The MTE as function of the perturbation amplitude. Shown are Monte-Carlo simulations (symbols), the theoretical result for the linear correction according to (\ref{eq:LT_ds_full_correction}) (solid line) and the unperturbed MTE [Eq.~(\ref{eq:ZeroEstablishmentRate})] (dashed horizontal line). Here $N=200$, $f_0=0.4$ and $\w=0.4$, and the theoretical linear correction is multiplied by a prefactor of 2.7 compared to the unperturbed MTE, to enable direct comparison with the simulations.}
	\label{fig5}
\end{figure}

\subsubsection{\textbf{Weak perturbation - Linear Theory}} \label{LT_section}
\begin{figure}[t]
	\includegraphics[width=0.81\linewidth]{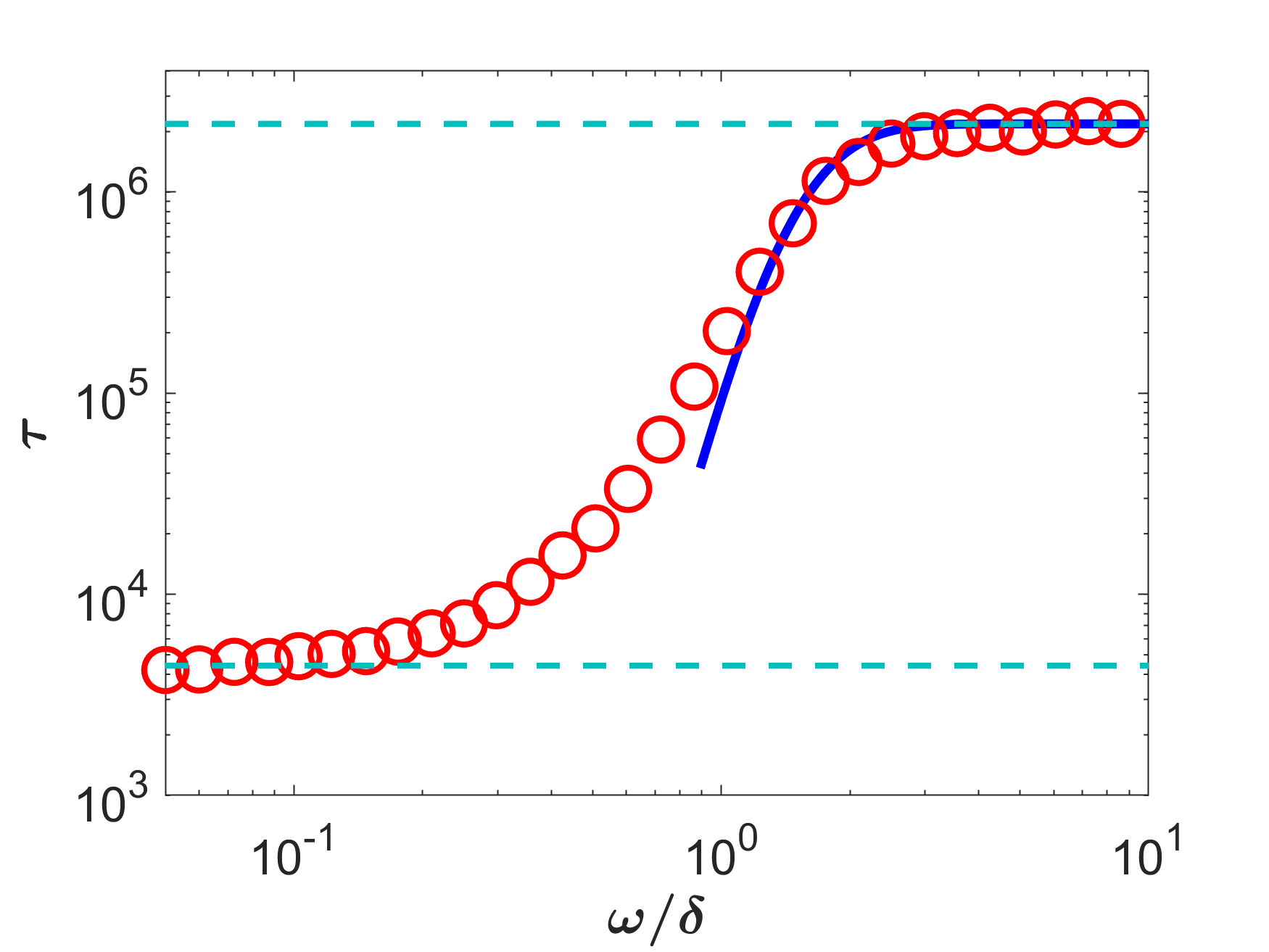}
	\caption{The MTE as function of the perturbation frequency $\w$ divided by $\delta=\sqrt{1-2f_0}$: Monte-Carlo simulations (symbols), theoretical expressions for the adiabatic regime according to~(\ref{eq:adiabaticRateFull}) (lower dashed line) and for the unperturbed MTE~(\ref{eq:ZeroEstablishmentRate}) (upper dashed line), and linear theory correction according to (\ref{eq:LT_ds_full_correction}) (solid line). Here $N=250$, $f_0=0.42$ and $\e=0.12$. The deviation of the linear theory from the simulation results stems from the fact the the pre-exponent depends also on $\w$.}
	\label{fig6}
\end{figure}

Here we assume that the perturbation amplitude is small, $\e \ll 1$. In this limit, it can be shown that the action (\ref{eq:Sperturbed}) satisfies $S(t_0) = S_0 + \Delta S(t_0)$~\cite{dykman1997resonant,dykman2001activated,assaf2008population,escudero2008persistence}, where $S_0$ is the action along the unperturbed optimal path $\{q_0(t-t_0), p_0(t-t_0)\}$, see Eqs.~(\ref{eq:ZeroTDtajectories}) and (\ref{eq:ZeroAction}), while  $\Delta S$ is the correction to action, given by~\cite{assaf2008population,escudero2008persistence}. Using Eq.~(\ref{eq:H1}), the latter satisfies
\begin{eqnarray} \label{eq:periodic_action}
\Delta S(t_0) &=& -\varepsilon\intop_{-\infty}^{\infty} \HH_1[q_0(t-t_0),p_0(t-t_0),t]dt\\
&=&-\e \frac{1-\delta^2}{2} \intop_{-\infty}^{\infty} \left[ \frac{2\sigma(t-t_0)}{1-\delta^2 + \sigma^2 (t-t_0)} -1 \right]\sin(\w t) dt,\nonumber
\end{eqnarray}
where $\sigma(t)$ was defined in Eq.~(\ref{eq:ZeroTDtajectories}). Solving this integral yields
\begin{eqnarray} \label{eq:dS_LT_pre}
\Delta S(t_0) &=&
-\e\pi\sqrt{1-\delta^2}
\csch\left(\frac{\pi\w}{\delta}\right) \sinh\left[ \frac{\w}{\delta} \arcsin(\delta) \right] \nonumber\\
&\times& \sin\left[\w t_0-\frac{\w}{\delta} \arctanh(\delta)\right],
\end{eqnarray}
where to remind the reader, $\delta = \sqrt{1-2f_0}$, and $\csch(x)=1/\sinh(x)$. Evaluating $\Delta S(t_0)$ at its minimum, $t_0^*(\w,\delta)=\pi/(2\w)+\arctanh(\delta)/\delta$, yields the linear correction to action:
\begin{equation} \label{eq:LT_ds_full_correction}
\Delta S = -\e\pi\sqrt{1-\delta^2}\csch\left(\frac{\pi\w}{\delta}\right)\sinh\left[ \frac{\w}{\delta} \arcsin(\delta) \right].
\end{equation}
Since the total action linearly decreases with $\e$, we find that the periodic perturbation exponentially decreases the MTE. This behavior can be seen in Fig.~\ref{fig5}, where we plot our analytical result, $\tau\sim e^{N(S_0+\Delta S)}$, as function of $\e$, with $\Delta S$ given by Eq.~(\ref{eq:LT_ds_full_correction}), along with numerical Monte-Carlo simulations. Additionally, in Fig.~\ref{fig6} we plot the MTE as a function of the perturbation frequency for small $\e$. In both figures we obtain good agreement between analytical and numerical results, as long as the perturbation amplitude is small.

It is interesting to look at Eq.~(\ref{eq:LT_ds_full_correction}) in the opposite limits of slowly- and rapidly-varying perturbation. In the former, the adiabatic limit, $\w \ll \delta$, we find
\begin{eqnarray} \label{eq:LT_ds_adiabatic_correction}
\hspace{-4mm}\Delta S = -\e \sqrt{1\!-\!\delta^2} \arcsin(\delta)
\left\{1\!-\!\frac{\w^2}{6\delta^2} \left[\pi^2\!-\!\arcsin^2(\delta)\right]\right\}.
\end{eqnarray}
From this expression it is evident that the leading-order term is constant with respect to $\w$, see Fig.~\ref{fig6} and the next subsection. Additionally, the fact that $\Delta S$ has an $\OO(\w^2)$ correction in the adiabatic limit is a generic feature, which has also been found in Ref.~\cite{assaf2008population} in a different model. 

In contrast, the limit of rapidly-varying perturbation, $\w\gg\delta$, yields
$\Delta S = -\pi\e\sqrt{1-\delta^2} \exp\{-(\w/\delta) \left[\pi-\arcsin(\delta)\right]\}$,
which decays exponentially as $\w$ grows. Thus, the linear (in $\e$) contribution vanishes in the limit of large $\w$. This indicates that in the limit of large $\w$, $\Delta S$ scales as some higher power of $\e$. In fact it has been shown in other models, see Refs.~\cite{assaf2008population,vilk2018population}, that for $\w \gg t_r^{-1}$, the correction to action scales in the leading order as $\OO(\e^2)$.

\subsubsection{\textbf{Adiabatic Approximation}} \label{Adiabatic}
We now compute the MTE in the adiabatic limit, $w \ll \delta$, using a different approach, which allows finding the pre-exponential correction to the MTE as well. In this limit, the mean \textit{rate} of establishment, $\bar{r}_{es}$, reads~\cite{assaf2008population}
\begin{equation}\label{ravg}
\bar{r}_{\text{es}}=\frac{\w}{2\pi} \intop_0^{2\pi/\w} r_{\text{es}}(t')dt',
\end{equation}
where $r_{\text{es}}(t)$ denotes the \textit{instantaneous} establishment rate. Since in this limit the external perturbation changes slowly, we refer to time in Hamiltonian~(\ref{eq:H0plusH1}) as a parameter, which allows finding the instantaneous optimal path and action. The resulting instantaneous establishment rate is given by
\begin{equation} \label{eq:adiabaticRate}
r_{\text{es}}(t) = \frac{\kk(t)\sqrt{1-\kk(t)}}{2\pi\sqrt{1+\kk(t)}}
e^{-N S(t)},
\end{equation}
where $S(t) = 2\kk(t) - 2\sqrt{1-\kk(t)^2}\arcsin\left[\kk(t)\right]$, and $\kk(t) = \sqrt{\delta^2-(1-\delta^2)\e\sin(\w t)}$. This result coincides with $\tau^{-1}$ from Eq.~(\ref{eq:ZeroEstablishmentRate})] upon replacing $\delta$ with $\kk(t)$.

As $r_{\text{es}}(t)$ receives its maximum at $t_* = \pi/(2\w)$, defining $\kappa_*\equiv\kk(t_*) = \sqrt{\delta^2-(1-\delta^2)\e}$, plugging~(\ref{eq:adiabaticRate}) into~(\ref{ravg}) and using the saddle point approximation around $t_*$, we obtain:
\begin{eqnarray} \label{eq:adiabaticRateFull}
\bar{r}_{es} &=& \frac{(1+\e)^{1/2}\kk_* (1-\kk_*)^{1/4}}{(2\pi)^{3/2}[N \e \arcsin(\kk_*)]^{1/2} (1+\kk_*)^{3/4}}
\nonumber\\
&\times&
e^{-N \left[2\kk_*-2\sqrt{1-\kk_*^2} \arcsin(\kk_*)\right]}.
\end{eqnarray}
Note that the  argument in the exponent is (minus) $N$ times the unperturbed action $S_0$, given by Eq.~(\ref{eq:ZeroAction}), upon replacing $\delta$ by $\kappa_*$. This means that in the adiabatic limit the system tends to switch to the established state when the environmental perturbation is at its maximal value.

\begin{figure}[t]
	\centering
	\includegraphics[width=0.95\linewidth]{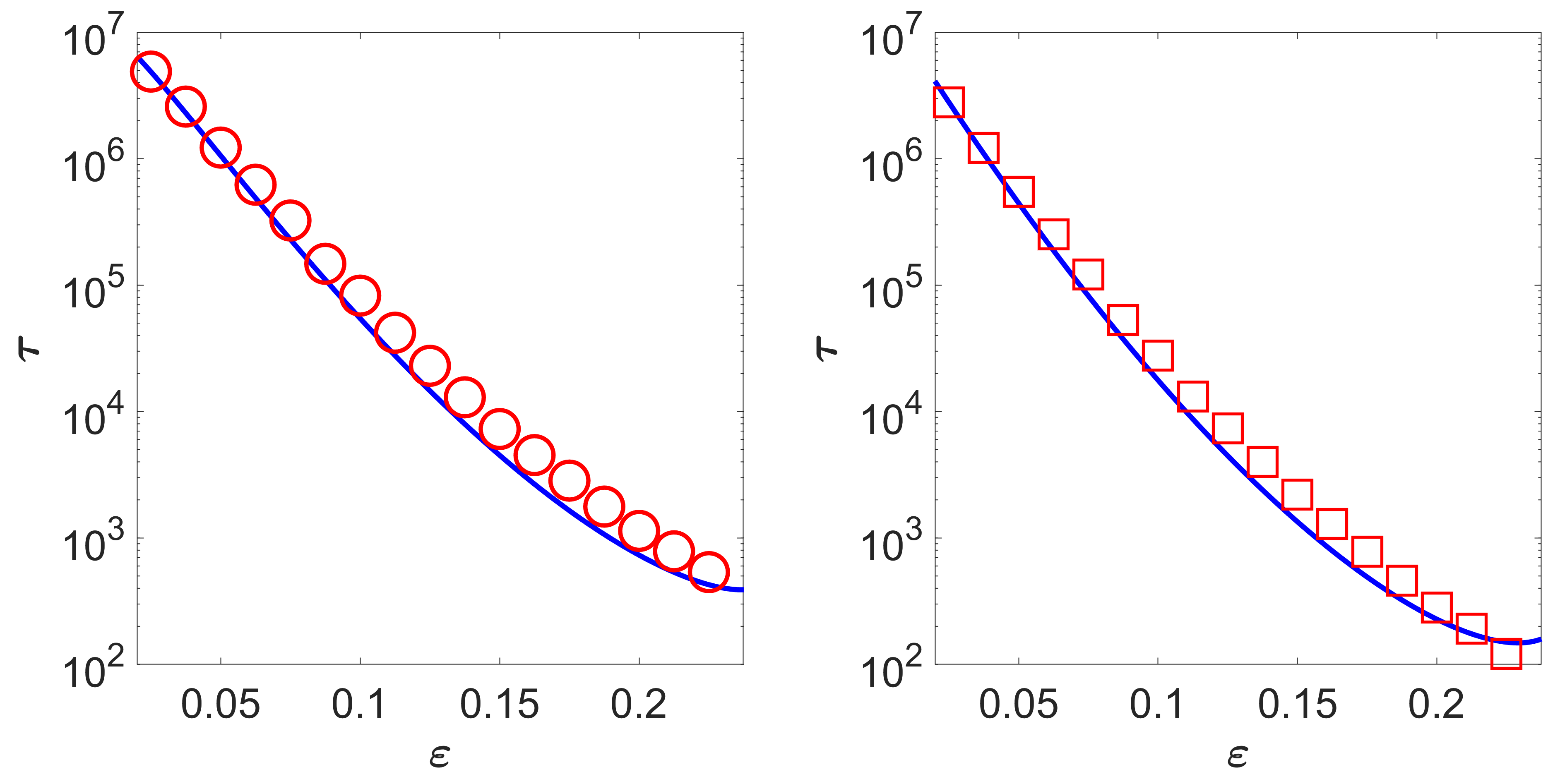}
	\caption{The MTE as function of the perturbation amplitude in the adiabatic limit. Left panel: sinusoidal perturbation -- the theoretical expression for the MTE~(\ref{eq:adiabaticRateFull}) with $\tau=1/\bar{r}_{es}$ (solid line), and numerical Monte-Carlo simulations (circles). Right panel: square wave perturbation -- the theoretical expression for the MTE~(\ref{eq:adiabaticRateSquare}) with $\tau=1/\bar{r}_{es}$ (solid line), and numerical Monte-Carlo simulations (squares). In both panels the parameters are $N=200$, $f_0=0.4$ and $\w=0.02$. One can see that as $\e$ is increased toward $\e_c$ (here $\e_c=0.25$), the theoretical result becomes invalid [see Eq.~(\ref{eps})], since the action approaches zero.}
	\label{fig7}
\end{figure}

The results of the adiabatic limit are shown in Fig.~\ref{fig7} where the analytical expression for the MTE, $\tau=1/\bar{r}_{\text{es}}$, with $\bar{r}_{es}$ given by Eq.~(\ref{eq:adiabaticRateFull}), is compared with numerical Monte-Carlo simulations.
Additionally, for a specific choice of perturbation amplitude $\e$, the MTE remains constant with respect to the perturbation frequency $\w$, as demonstrated by Eq.~(\ref{eq:adiabaticRateFull}) and supported by simulation results shown in Fig.~\ref{fig6}.

To check the consistency of our adiabatic approximation, we can expand the logarithm of the mean rate of establishment up to first order in $1/N \ll \e \ll 1$. Doing so, we obtain:
\begin{equation} \label{eq:AdiabaticSinRate}
NS \simeq -\ln(\bar{r}_{\text{es}}) \simeq
N S_0
-\e N\sqrt{1-\delta ^2}\arcsin(\delta),
\end{equation}
where $S_0$ is given by~(\ref{eq:ZeroAction}). One can see that the linear term in $\e$ agrees with $\Delta S$ [Eq.~(\ref{eq:LT_ds_adiabatic_correction})] in the leading order in $\omega\ll \delta$.

Note that the adiabatic mean rate of establishment~(\ref{eq:adiabaticRateFull}) has a maximum cutoff value at $\e_c = \delta^2/(1-\delta^2)$  (for which $\kappa_*=0$) beyond which the action becomes negative, and therefore the rate has no physical meaning.
However, since the WKB approximation requires that the action be large, it breaks down well before $\e$ reaches $\e_c$. To find the exact condition of applicability of the WKB approximation, we expand the argument in the exponent of Eq.~(\ref{eq:adiabaticRateFull}) around $\kk_*=0$, yielding
$2N [\kk_*-(1-\kk_*^2)^{-1/2} \arcsin(\kk_*)]
=(2/3) N \kk_*^3 + \OO (\kk_*^5)$.
As a result, since $\kappa_*\sim (\e_c-\e)^{1/2}$, the WKB approximation is applicable as long as $N\kappa_*^3 \gg 1$, or
\begin{equation}\label{eps}
\e_c - \e \gg N^{-2/3}.
\end{equation}
The fact that when approaching $\e_c$, theoretical predictions~(\ref{eq:adiabaticRateFull}) becomes invalid, is demonstrated in Fig.~\ref{fig7}.

Finally, one can also study other time-periodic perturbations. As an additional example, we consider a square-wave perturbation, such that Eq.~(\ref{eq:sinusoidal_pert}) becomes $f(t) = f_0\left\{1+\e \sign\left[\sin(\w t)\right]\right\}$, see Fig.~\ref{fig1}. A similar treatment in this case yields the mean rate of establishment
\begin{equation} \label{eq:adiabaticRateSquare}
\bar{r}_{es} = \frac{\kk_*\sqrt{1-\kk_*}}{4\pi\sqrt{1+\kk_*}} e^{-N \left[2\kk_*-2\sqrt{1-\kk_*^2} \arcsin(\kk_*)\right]}.
\end{equation}
Note that the argument in the exponent of Eq.~(\ref{eq:adiabaticRateSquare}) coincides with that of the sinusoidal case, Eq.~(\ref{eq:adiabaticRateFull}). This indicates that the parameter that controls the MTE in the adiabatic case is the perturbation amplitude $\e$, while the exact shape of the periodic perturbation is unimportant in the leading order in $\omega\ll\delta$. In Fig.~\ref{fig7} we compare the theoretical result, Eq.~(\ref{eq:adiabaticRateSquare}), with numerical Monte-Carlo simulations, and find very good agreement, as long as $\e$ is not too close to $\e_c$.

\section{Simulation methods} \label{SimulationMethods}
In this section we briefly describe our numerical simulations and the algorithm behind the modified Gillespie algorithm \cite{gillespie1976general} with time-dependent rates. 

Throughout this work, to compute switching probabilities under a temporal environmental perturbation, we ran many realizations until a given time $t_{end}$. Here, $t_0$ -- the onset of perturbation -- was taken to be several times the relaxation time of the system. Once the realization has reached $t_0+T$, we turned off the perturbation, waited several relaxation times and terminated the simulation. The switching probability was determined by the fraction of realizations that switched up to this designated time $t_{end}$ out of all realizations. To compute the mean switching time we averaged over the switching times of 1000 simulations for each parameter set, where each simulation terminated when the system crossed the threshold for switching. In all our simulations, for each set of parameters we made sure that the numerical error, which is on the order of $1/\sqrt{\mathcal{N}}$ with $\mathcal{N}$ being the number of completed simulations, was at most $10\%$. In all figures, symbol sizes represent the maximal error of the simulations.

To implement a Gillespie algorithm with explicitly time-dependent reaction rates we had to properly sample the time until the next reaction. For time-\textit{independent} reaction rates, the time until the next reaction, $\Delta$, is exponentially distributed, with mean that equals the inverse of the sum of the rates. Yet, for explicitly time-dependent rates, $\lambda(n,t)$ and $\mu(n,t)$, the distribution of $\Delta$ can be shown to satisfy~\cite{anderson2007modified}: $P(\Delta) = 1-\exp\{-\intop_t^{t+\Delta} [\lambda(n(t),s)+\mu(n(t),s)]ds\}$.
As a result, $\Delta$ can be found by solving the equation
\begin{equation} \label{eq:reactionsGillespie}
\intop_t^{t+\Delta} \left\{\lambda\left[n(t),s\right]+\mu\left[n(t),s\right]\right\}ds
= \ln(1/r),
\end{equation}
where $r$ is a random number drawn from a uniform distribution $U(0,1)$.

Since solving Eq.~(\ref{eq:reactionsGillespie}) for $\Delta$ at each time step is very time consuming, to improve the efficiency of our simulation, we have constructed a 3D matrix, in which $\Delta$ is computed for each combination $(n,t,r)$ \textit{before} the initialization of the simulation. Then, at each time step, we find the closest matrix element which matches the current values of $n$, $t$ and $r$, and in this way, we are able to find $\Delta$ in an efficient manner.

To avoid too large a matrix, we note that, what determines the perturbation magnitude (in both cases we have considered) is not the absolute time, but the time relative to the last period, or the time since the perturbation onset (in the temporary change case). We have verified that our results were converged when taking matrix of size $\ceil{n_2}\times40\times1000$; that is, the relative time has been divided into 40 bins, and the logarithm of $r$ [where $r\in(0,1)$ is a random uniform number] into a 1000 bins. We have checked that increasing the number of time intervals beyond 40 and the number of $r$ intervals beyond 1000 had a negligible effect on the results.

\begin{figure}[t]
	\includegraphics[width=0.81\linewidth]{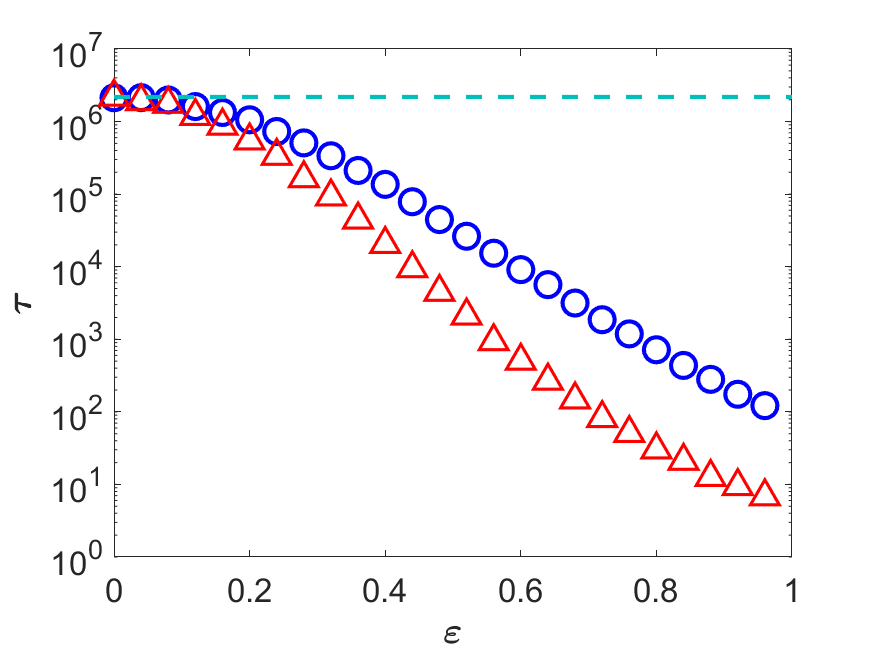}
	\caption{The MTE versus the sinusoidal perturbation amplitude $\varepsilon$. Monte-Carlo simulations for the additive case, \textit{i.e.} with Eq.~(\ref{eq:sinusoidal_pert}) (circles), compared to the multiplicative case, using Eq.~(\ref{eq:reactionsMultiplicativeConst}) with $\phi(t)=f_0 \e \sin(\w t)$ (triangles). For $\e\to 0$, both cases converge to the unperturbed MTE (dashed horizontal line), given by Eq.~(\ref{eq:ZeroEstablishmentRate}). Parameters are $N=250$, $f_0=0.42$ and $\w=1$. For $\e\ll 1$, the effects are comparable. Yet, a higher amplitude results in a greater effect in the multiplicative case, see text.}
	\label{fig8}
\end{figure}

\section{Discussion and Conclusions} \label{summary}
In this work we have studied population switching under a time-varying environment. Our underlying model included a stochastic population that undergoes noise-driven switching, from one long-lived metastable state, called the pre-established state, into another, called the established state.

We have found that the establishment probability (EP) and the mean time for establishment (MTE) of such populations are strongly affected by a time-varying environment and its characteristics. In the case of a temporary change in the environment, we have developed a generic framework that allowed computing the increase in the EP as a function of the perturbation magnitude and duration. We further found the EP close to the bifurcation limit, in which the result was drastically simplified.

In the case of a periodically-varying environment, we have obtained analytical results in the limit of weak perturbation, and also in the adiabatic regime, where the perturbation frequency is small compared to the typical relaxation rate of the system. Here, in addition to a sinusoidal perturbation, we have also considered a square wave perturbation. By doing so, we have demonstrated that in the adiabatic regime, the MTE is governed in the leading order by the perturbation amplitude rather than the specific shape of the perturbation.

We have also checked the effect of a multiplicative periodic perturbation on the MTE compared to the additive one. In order to do so, we ran Monte-Carlo simulations with reaction rates given by Eqs.~(\ref{eq:reactionsAdditive}) and~(\ref{eq:reactionsMultiplicativeConst}), both with $\phi(t)=f_0\e \sin(\w t)$. In Fig.~\ref{fig8} we show simulation results for the MTE as function of the perturbation amplitude comparing the additive and multiplicative cases. One can see that for a small perturbation amplitude $\e\ll1$ the effects are comparable; however for $\e=\OO(1)$ the MTE in the multiplicative case is significantly shorter than the additive one. This behavior stems from the fact that in the multiplicative case, the death rate vanishes as the perturbation amplitude approaches 1. This qualitative behavior was also observed when simulating the case of a temporary perturbation.

Finally, the analytical approach that we have developed here can be useful to analyze genetic circuits which display phenotypic switching. Since such genetic switches often operate in a noisy environment, the switching time is expected to strongly depend on the characteristics of the environmental variations, such as the variation magnitude, and its frequency or duration, see \textit{e.g.}~\cite{elowitz2002stochastic,assaf2013extrinsic,volfson2006origins}. Being able to theoretically assess how such environmental variability affects the switching time in such genetic circuits, comparing these predictions to experimental results may allow us to infer key biological parameters, and even epigenetic landscapes, in such complex gene regulatory networks.

We acknowledge support from the Israel Science Foundation grant No. 300/14 and the United States-Israel Binational Science Foundation grant
No. 2016-655.

\bibliographystyle{apsrev4-1}
\bibliography{IA2019_bib}

\end{document}